\documentclass{article}
\usepackage{amsmath,graphicx,url,amssymb,latexsym,epsfig,tabularx,setspace,verbatim,rotating,threeparttable}
\usepackage{natbib}
\usepackage[hmargin=1.5in,vmargin=1.5in]{geometry}

\begin{document}
\large

\doublespace

\voffset 0.8in
\setcounter{secnumdepth}{4}
\begin{titlepage}
\centerline{\bf \LARGE The Habitable Epoch of the Early Universe}

\vskip 0.7cm

\large
\centerline{\Large Abraham Loeb}
\vspace{0.5cm}
\large

\noindent
{\it Astronomy Department, Harvard University, 60 Garden Street, 
Cambridge, MA 02138, USA; {\it E-mail:} aloeb@cfa.harvard.edu}\\ 



\vskip 1cm

\large

\begin{abstract}
\large
 
In the redshift range $100\lesssim (1+z)\lesssim 137$, the cosmic
microwave background (CMB) had a temperature of 273--373 K
(0-100$^\circ$C), allowing early rocky planets (if any existed) to
have liquid water chemistry on their surface and be habitable,
irrespective of their distance from a star. In the standard
$\Lambda$CDM cosmology, the first star-forming halos within our Hubble
volume started collapsing at these redshifts, allowing the chemistry
of life to possibly begin when the Universe was merely 10--17 million
years old. The possibility of life starting when the average matter
density was a million times bigger than it is today argues against the
anthropic explanation for the low value of the cosmological constant.

\end{abstract}

\vskip 0.2in

\large
\noindent
{\bf Kewords:} first stars, habitability, cosmology

\end{titlepage}

\large

\section{Introduction}

The habitable zone is commonly defined in reference to a distance from
a luminous source, such as a star \citep{Kasting, Kasting1}, whose
heat maintains the surface of a rocky planet at a temperature of $\sim
300$K, allowing liquid water to exist and the chemistry of ``life as
we know it'' to operate.  In this brief paper, I point out that the
cosmic microwave background (CMB) provided a uniform heating source at
a temperature of $T_{\rm cmb}=272.6 {\rm K} \times [(1+z)/100]$
\citep{Fixsen} that could have made by itself rocky planets habitable
at redshifts $(1+z)=100$--137 in the early Universe, merely 10--17
million years after the Big Bang.

In order for rocky planets to exist at these early times, massive
stars with tens to hundreds of solar masses, whose lifetime is much
shorter than the age of the Universe, had to form and enrich the
primordial gas with heavy elements through winds and supernova
explosions \citep{Ober,Heger}. Indeed, numerical simulations predict
that predominantly massive stars have formed in the first halos of
dark matter to collapse \citep{BL04,LF}.  For massive stars that are
dominated by radiation pressure and shine near their Eddington
luminosity $L_{\rm E}=1.3\times 10^{40}~{\rm
erg~s^{-1}}(M_\star/100M_\odot)$, the lifetime is independent of
stellar mass $M_\star$ and set by the 0.7\% nuclear efficiency for
converting rest mass to radiation, $\sim (0.007M_\star c^2)/L_{\rm E}=
3~{\rm Myr}$ \citep{El,BKL}. We next examine how early did such stars
form within the observable volume of our Universe.

\section{First Planets}

In the standard cosmological model, structure forms hierarchically --
starting from small spatial scales, through the gravitational growth
of primordial density perturbations \citep{LF}. On any given spatial
scale $R$, the probability distribution of fractional density
fluctuations $\delta$ is assumed to have a Gaussian form,
$P(\delta)d\delta
=(2\pi\sigma^2)^{-1/2}\exp\{-\delta^2/2\sigma^2\}d\delta$, with a
{\it root-mean-square} amplitude $\sigma(R)$ that is initially much
smaller than unity.  The initial $\sigma(R)$ is tightly constrained on
large scales, $R\gtrsim 1~{\rm Mpc}$, through observations of the CMB
anisotropies and galaxy surveys \citep{Planck,And}, and is
extrapolated theoretically to smaller scales. Throughout the paper, we
normalize spatial scales to their so-called ``comoving'' values in the
present-day Universe. The assumed Gaussian shape of $P(\delta)$ has so
far been tested only on scales $R\gtrsim 1~{\rm Mpc}$ for $\delta
\lesssim 3\sigma$ \citep{Shan}, but was not verified in the far tail
of the distribution or on small scales that are first to collapse in
the early Universe.

As the density in a given region rises above the background level, the
matter in it detaches from the Hubble expansion and eventually
collapses owing to its self-gravity to make a gravitationally bound
(virialized) object like a galaxy. The abundance of regions that
collapse and reach virial equilibrium at any given time depends
sensitively on both $P(\delta)$ and $\sigma(R)$. Each collapsing
region includes a mix of dark matter and ordinary matter (often
labeled as ``baryonic'').  If the baryonic gas is able to cool below
the virial temperature inside the dark matter halo, then it could
fragment into dense clumps and make stars.

At redshifts $z\gtrsim 140$ Compton cooling on the CMB is effective on
a timescale comparable to the age of the Universe, given the residual
fraction of free electrons left over from cosmological recombination
(see \S 2.2 in \cite{LF} and \cite{PL12}). The thermal coupling to the
CMB tends to bring the gas temperature to $T_{\rm cmb}$, which at
$z\sim 140$ is similar to the temperature floor associated with
molecular hydrogen cooling \citep{Haiman,Tegmark,Hirata}. In order for
virialized gas in a dark matter halo to cool, condense and fragment
into stars, the halo virial temperature $T_{\rm vir}$ has to exceed
$T_{\rm min}\approx 300$K, corresponding to $T_{\rm cmb}$ at
$(1+z)\sim 110$. This implies a halo mass in excess of $M_{\rm
min}=10^4 M_\odot$, corresponding to a baryonic mass $M_{\rm b,
min}=1.5\times 10^3~M_\odot$, a circular virial velocity $V_{\rm c,
min}=2.6~{\rm km~s^{-1}}$ and a virial radius $r_{\rm vir,min
}=6.3~{\rm pc}$ (see \S 3.3 in \cite{LF}). This value of $M_{\rm min}$
is close to the minimum halo mass to assemble baryons at that redshift
(see \S 3.2.1 in \cite{LF} and Fig. 2 of \cite{TBH}).

The corresponding number of star-forming halos on our past light cone
is given by \citep{Naoz},
\begin{equation}
N=\int_{(1+z)=100}^{(1+z)=137} n(M>M_{\rm min},z^\prime) {dV\over
dz^\prime} dz^\prime ,
\end{equation}
where $n(M>M_{\rm min})$ is the comoving number density of halos with
a mass $M>M_{\rm min}$ \citep{Sheth}, and $dV=4\pi r^2dr$ is the
comoving volume element with $dr=cdt/a(t)$. Here, $a(t)=(1+z)^{-1}$ is
the cosmological scale factor at time $t$, and $r(z)=c\int_0^z
dz^\prime/H(z^\prime)$ is the comoving distance. The Hubble parameter
for a flat Universe is $H(z)\equiv (\dot{a}/a)
=H_0\sqrt{\Omega_m(1+z)^3+\Omega_r(1+z)^4+\Omega_\Lambda}$, with
$\Omega_m$, $\Omega_r$ and $\Omega_\Lambda$ being the present-day
density parameters of matter, radiation and vacuum, respectively. The
total number of halos that existed at $(1+z)\sim 100$ within our
entire Hubble volume (not restricted to the light cone), $N_{\rm
tot}\equiv n (M>M_{\rm min},z=99)\times (4\pi/3)(3c/H_0)^3$, is larger
than $N$ by a factor of $\sim 10^3$.

For the standard cosmological parameters \citep{Planck}, we find that
the first star-forming halos on our past light cone reached its
maximum turnaround radius\footnote{In the spherical collapse model,
the turnaround time is half the collapse time.} (with a density
contrast of 5.6) at $z\sim 112$ and collapsed (with an average density
contrast of 178) at $z\sim 71$. Within the entire Hubble volume, a
turnaround at $z\sim 122$ resulted in the first collapse at $z\sim
77$.  This result includes the delay by $\Delta z\sim 5.3$ expected
from the streaming motion of baryons relative to the dark matter
\citep{Fialkov}.

The above calculation implies that rocky planets could have formed
within our Hubble volume by $(1+z)\sim 78$ but not by $(1+z)\sim 110$
if the initial density perturbations were perfectly Gaussian. However,
the host halos of the first planets are extremely rare, representing
just $\sim 2\times 10^{-17}$ of the cosmic matter inventory. Since
they lie $\sim 8.5$ standard deviations ($\sigma$) away on the
exponential tail of the Gaussian probability distribution of initial
density perturbations, $P(\delta)$, their abundance could have been
significantly enhanced by primordial non-Gaussianity
\citep{LV,Maio,Musso} if the decline of $P(\delta)$ at high values of
$\delta/\sigma$ is shallower than exponential. The needed level of
deviation from Gaussianity is not ruled out by existing data sets
\citep{Planck2}. Non-Gaussianity below the current limits is expected
in generic models of cosmic inflation \citep{Malda} that are commonly
used to explain the initial density perturbations in the Universe.

\section{Discussion}

In this brief paper, I highlighted a new regime of habitability made
possible for $\sim 6.6$ Myr by the uniform CMB radiation at redshifts
$(1+z)= 100$--137, just when the first generation of star-forming
halos (with a virial mass $\gtrsim 10^4M_\odot$) turned around in the
standard cosmological model with Gaussian initial
conditions. Deviations from Gaussianity in the far ($8.5\sigma$) tail
of the probability distribution of initial density perturbations,
could have led already at these redshifts to the birth of massive
stars, whose heavy elements triggered the formation of rocky planets
with liquid water on their surface.\footnote{The dynamical time of
galaxies is shorter than $\sim 1/\sqrt{200}= 7\%$ of the age of the
Universe at any redshift since their average density contrast is
$\gtrsim 200$. After the first stars formed, the subsequent delay in
producing heavy elements from the first supernovae could have been as
short as a few Myr. The supernova ejecta could have produced
high-metallicity islands that were not fully mixed with the surrounding
primordial gas, leading to efficient formation of rocky planets within
them.}

Thermal gradients are needed for life. These can be supplied by
geological variations on the surface of rocky planets. Examples for
sources of free energy are geothermal energy powered by the planet's
gravitational binding energy at formation and radioactive energy from
unstable elements produced by the earliest supernova. These internal
heat sources (in addition to possible heating by a nearby star), may
have kept planets warm even without the CMB, extending the habitable
epoch from $z\sim 100$ to later times.  The lower CMB temperature at
late times may have allowed ice to form on objects that delivered
water to a planet's surface, and helped to maintain the cold trap of
water in the planet's stratosphere.  Planets could have kept a blanket
of molecular hydrogen that maintained their warmth
\citep{Stevenson,Gaidos}, allowing life to persist on internally
warmed planets at late cosmic times. If life persisted at $z\lesssim
100$, it could have been transported to newly formed objects through
panspermia \citep{Nichol}. Under the assumption that interstellar
panspermia is plausible, the redshift of $z\sim 100$ can be regarded
as the earliest cosmic epoch after which life was possible in our
Universe.

The feasibility of life in the early universe can be tested by
searching for planets with atmospheric bio-signatures around
low-metallicity stars in the Milky Way galaxy or its dwarf galaxy
satellites. Such stars represent the closest analogs to the first
generation of stars at early cosmic times.
 
The possibility that the chemistry of life could have started in our
universe only 10--17 Myr after the Big Bang argues against the
anthropic explanation\footnote{In difference from \cite{Weinberg89},
we require here that stars form in any low-mass halo rather than in a
galaxy as massive as the Milky-Way, as the pre-requisite for life.}
for the value of the cosmological constant \citep{Weinberg89},
especially if the characteristic amplitude of initial density
perturbations or the level of non-Gaussianity is allowed to vary in
different regions of the multiverse\footnote{An increase in the
initial amplitude of density perturbations on the mass scale of
$10^4M_\odot$ by a modest factor of $1.4\times [(1+z)/110]$ would have
enabled star formation within the Hubble volume at redshifts
$(1+z)>110$ even for perfectly Gaussian initial conditions.}
\citep{Vilenkin,Tegmark1}. In principle, the habitable cosmological
epoch considered here allows for life to emerge in a Universe with a
cosmological constant that is $(1+z)^3\sim 10^6$ times bigger than
observed \citep{Loeb05}. If observers can eventually emerge from
primitive forms of life at an arbitrarily later time in such a
Universe, then their existence would be in conflict with the anthropic
reasoning for the low value of the cosmological constant in our
Universe. Even when placed on a logarithmic scale, the corresponding
discrepancy in the vacuum energy density is substantial, spanning
$\sim 5\%$ of the $\sim 120$ orders of magnitude available up to the
Planck density. The volume associated with inflating regions of larger
vacuum density is exponentially greater than our region, making
residence in them far more likely.

\vskip 0.65in
\noindent
{\bf ACKNOWLEDGEMENTS.} I thank F. Dyson, J. Maldacena, D. Maoz,
E. Turner for useful comments on the manuscript. This work was
supported in part by NSF grant AST-1312034.

\newpage


\bibliographystyle{apj}


\end{document}